\documentclass[conference]{IEEEtran}
\usepackage{graphicx}
\usepackage{amssymb}
\usepackage[space]{cite}
\usepackage{multirow}
\usepackage[acronym,nomain]{glossaries}
\usepackage{booktabs}
\usepackage{color,soul}
\usepackage{threeparttable}
\usepackage[per-mode=symbol]{siunitx}
\usepackage[inkscapelatex=false]{svg}
\usepackage{balance}
\usepackage{array}
\makeglossaries
\newacronym{2d}{2D}{two-dimensional}
\newacronym{3d}{3D}{three-dimensional}
\newacronym{5g}{5G}{5th-generation}
\newacronym{pnt}{PNT}{positioning, navigation and timing}
\newacronym{pa}{PA}{physical anchor}
\newacronym{va}{VA}{virtual anchor}
\newacronym{bs}{BS}{base station}
\newacronym{mpc}{MPC}{multipath component}
\newacronym{slam}{SLAM}{simultaneous localization and mapping}
\newacronym{ue}{UE}{mobile user}
\newacronym{fa}{FA}{false alarm}

\newacronym{adc}{ADC}{analog-to-digital converter}
\newacronym{ai}{AI}{Artificial Intelligence}
\newacronym{aoa}{AoA}{angle-of-arrival}
\newacronym{aod}{AoD}{angle-of-departure}
\newacronym{ap}{AP}{access point}
\newacronym{asic}{ASIC}{application specific integrated circuit}
\newacronym{asip}{ASIP}{application specific integrated processor}
\newacronym{awgn}{AWGN}{additive white Gaussian noise}

\newacronym{bp}{BP}{belief propagation}

\newacronym{ce}{CE}{channel estimation}
\newacronym{cw}{CW}{continuous wave}
\newacronym{cdf}{CDF}{cumulative distribution function}
\newacronym{ced}{CED}{cumulative energy density}
\newacronym{clk}{CLK}{clock}
\newacronym{csi}{CSI}{channel state information}
\newacronym{csp}{CSP}{contact service point}
\newacronym{cpu}{CPU}{central processing unit}
\newacronym{crlb}{CRLB}{Cramer-Rao lower bound}
\newacronym{dc}{DC}{direct current}
\newacronym{dfg}{DFG}{data-flow graph}
\newacronym{dlt}{DLT}{Distributed Ledger Technology}
\newacronym{dmimo}{D-MIMO}{distributed MIMO}
\newacronym{ram}{RAM}{random-access memory}
\newacronym{ecc}{ECC}{elliptic curve cryptography}
\newacronym{ecdlp}{ECDLP}{elliptic curve discrete logarithm problem}
\newacronym{ecsp}{ECSP}{edge computing service point}
\newacronym{eirp}{EIRP}{equivalent isotropically radiated power}
\newacronym{em}{EM}{electromagnetic}
\newacronym{en}{EN}{energy neutral}
\newacronym{e2e}{E2E}{End-to-End}
\newacronym{erfc}{erfc}{complementary error function}
\newacronym{pe}{PE}{processing engine}
\newacronym{rtl}{RTL}{register transfer level}
\newacronym{hls}{HLS}{high-level synthesis}
\newacronym{fft}{FFT}{fast fourier transform}
\newacronym{fhss}{FHSS}{frequency hopping spread spectrum}
\newacronym{fl}{FL}{federated learning}
\newacronym{fpga}{FPGA}{field programmable gate array}
\newacronym{gdpr}{GDPR}{general data protection regulation}
\newacronym{ic}{IC}{integrated circuit}
\newacronym{ioe}{IoE}{Internet of Everything}
\newacronym{iot}{IoT}{Internet of Things}
\newacronym{id}{ID}{information decoding}
\newacronym{kpi}{KPI}{key performance indicator}
\newacronym{lca}{LCA}{life cycle assessment}
\newacronym{ls}{LS}{least squares}
\newacronym{lis}{LIS}{large intelligent surface}
\newacronym{liion}{Li-ion}{lithium-ion}
\newacronym{los}{LoS}{line-of-sight}
\newacronym{lut}{LUT}{look-up table}
\newacronym{nlos}{NLoS}{non-line-of-sight}
\newacronym{olos}{OLoS}{obstructed-line-of-sight}
\newacronym{lmo}{LMO}{lithium ion manganese oxide}
\newacronym{mf}{MF}{matched filter}
\newacronym{ml}{ML}{machine learning}
\newacronym{mmse}{MMSE}{minimum mean square error}
\newacronym{mmwave}{mmWave}{millimeter wave}
\newacronym{mimo}{MIMO}{multiple-input multiple-output}
\newacronym{miso}{MISO}{multiple-input single-output}

\newacronym{mrc}{MRC}{maximum ratio combining}
\newacronym{mrt}{MRT}{maximum ratio transmission}
\newacronym{ofdm}{OFDM}{orthogonal frequency-division multiplexing}
\newacronym{ota}{OTA}{over-the-air}
\newacronym{ii}{II}{Initiation interval}
\newacronym{nb}{NB}{narrowband}
\newacronym{nr}{NR}{New Radio}
\newacronym{pda}{PDA}{probabilistic data association}
\newacronym{pdp}{PDP}{power delay profile}
\newacronym{pdf}{PDF}{probability density function}
\newacronym{spa}{SPA}{sum-product algorithm}

\newacronym{per}{PER}{packet error rate}
\newacronym{pet}{PET}{privacy enhancing technology}
\newacronym{pki}{PKI}{public key infrastucture}
\newacronym{pc}{PC}{pilot count}
\newacronym{pnn}{PNN}{probabilistic neural network}
\newacronym{plhf}{PLHF}{pseudo-likelihood function}
\newacronym{qrrls}{QR-RLS}{QR decomposition based recursive least squares}
\newacronym{re}{RE}{radio element}
\newacronym{rf}{RF}{radio frequency}
\newacronym{rfid}{RFID}{radio frequency identification}
\newacronym{ris}{RIS}{reflective intelligent surface}%reconfigurable?
\newacronym{rmse}{RMSE}{root mean square error}
\newacronym{rls}{RLS}{recursive least squares}
\newacronym{rss}{RSS}{received signal strength}
\newacronym{rssi}{RSSI}{received signal strength indicator}
\newacronym{rw}{RW}{RadioWeaves}
\newacronym{sa}{SA}{synchronization anchor}
\newacronym{sdg}{SDG}{Sustainable Development Goal}
\newacronym{sdn}{SDN}{software-defined network}
\newacronym{se}{SE}{spectral efficiency}
\newacronym{sinr}{SINR}{signal-to-interference-plus-noise ratio}
\newacronym{siso}{SISO}{single-input single-output}
\newacronym{snr}{SNR}{signal-to-noise ratio}
\newacronym{srls}{SRLS}{standard recursive least squares}
\newacronym{svd}{SVD}{singular value decomposition}
\newacronym{sp}{SP}{service point}
\newacronym{sram}{SRAM}{static random-access memory}
\newacronym{tdd}{TDD}{time division duplexing}
\newacronym{tdoa}{TDOA}{time-difference-of-arrival}
\newacronym{toa}{TOA}{time-of-arrival}
\newacronym{uhf}{UHF}{ultra high frequency}
\newacronym{ula}{ULA}{uniform linear array}
\newacronym{ulp}{ULP}{uplink pilot}
\newacronym{ul}{ULP}{uplink}
\newacronym{uld}{ULD}{uplink data}
\newacronym{upa}{UPA}{uniform planar array}
\newacronym{ura}{URA}{uniform rectangular array}
\newacronym{uwb}{UWB}{ultrawideband}
\newacronym{wb}{WB}{wideband}
\newacronym{wpt}{WPT}{wireless power transfer}
\newacronym{zf}{ZF}{zero forcing}

\newacronym{lpt}{LPT}{laser power transfer}
\newacronym{rfpt}{RFPT}{radio frequency power transfer}
\newacronym{ipt}{IPT}{inductive power transfer}
\newacronym{cpt}{CPT}{capacitive power transfer}
\newacronym{apt}{APT}{acoustic power transfer}
\newacronym{cfo}{CFO}{carrier frequency offset}
\newacronym{lo}{LO}{local oscillator}
\newacronym{if}{IF}{intermediate frequency}
\newacronym{duc}{DUC}{digital up conversion}
\newacronym{ddc}{DDC}{digital down conversion}
\newacronym{dsp}{DSP}{digital signal processing}
\newacronym{vco}{VCO}{voltage-controlled oscillator}
\newacronym{pll}{PLL}{phase-locked loop}
\newacronym{rt}{RT}{ray-tracing}
\newacronym{lna}{LNA}{low noise amplifier}
\newacronym{dac}{DAC}{digital-to-analog converter}
\newacronym{de}{DE}{drain efficiency}
\newacronym{pae}{PAE}{power-added efficiency}
\newacronym{sar}{SAR}{specific absorption rate}
\newacronym{mppt}{MPPT}{maximum power point tracking}
\newacronym{isi}{ISI}{intersymbol interference}
\newacronym{pps}{PPS}{pulse per second}
\newacronym{ps}{PS}{processing system}
\newacronym{pl}{PL}{programmable logic}
\newacronym{icnirp}{ICNIRP}{International Commission on Non-Ionizing Radiation Protection}
\newacronym{fd}{FD}{Front-haul distance}
\newacronym{wd}{WD}{Wireless distance}
\newacronym{ntp}{NTP}{Network Time Protocol}
\newacronym{ptp}{PTP}{Precision Time Protocol}
\newacronym{wr}{WR}{White Rabbit}
\newacronym{mm2s}{MM2S}{memory mapped to stream}
\newacronym{s2mm}{S2MM}{stream to memory mapped}
\newacronym{ber}{BER}{bit error rate}
\newacronym{mu}{MU}{multiplication unit}
\newacronym{gp}{GP}{general purpose}
\newacronym{hp}{HP}{high performance}
\newacronym{ids}{IDS}{indoor dense space}
\newacronym{dma}{DMA}{direct memory access}
\newacronym{qam}{QAM}{Quadrature amplitude modulation}
\newacronym{qpsk}{QPSK}{Quadrature  phase shift keying}
\newacronym{sd}{SD}{sphere decoding}
\newacronym{xr}{XR}{extended reality}

\usepackage[printonlyused,withpage]{acronym}

\newcommand*{\sectionshift}{-0.2cm}

 \usepackage{tkz-euclide}
 \usetikzlibrary{arrows}
 \usepackage{setspace}
 \usepackage{bm}
 \usepackage{relsize}
 \usepackage{psfrag}
 \usepackage{dblfloatfix}
 \usepackage{tabularx}
\usepackage{soul}
\usepackage{float}
 \usepackage{subfiles}
 \usepackage{epstopdf}
 \usepackage{verbatim}
 \usepackage{ifthen}
 \usepackage{pgfplots}
 \usepackage{pgfplotstable}
 \pgfplotsset{compat=1.18}
 \usepackage{tikzscale}
 \usepackage{amsfonts}
 \usepackage{tikz-3dplot}
 \usepackage{graphicx}
\usepackage{subfig}
\usepackage{footnote}
\makesavenoteenv{tabular}
\makesavenoteenv{table}

\usepackage{tikz}
\usepgfplotslibrary{fillbetween}
\usetikzlibrary{backgrounds}
\tikzstyle{load}   = [ultra thick,-latex]
\tikzstyle{stress} = [-latex]
\tikzstyle{dim}    = [latex-latex]
\tikzstyle{axis}   = [-latex,black!55]
\usetikzlibrary{arrows}
\usepackage{tikzscale}
\usepackage{tikz-3dplot}
\usetikzlibrary{3d}
\usetikzlibrary{shapes}
\usetikzlibrary {patterns,patterns.meta}
\usetikzlibrary{shapes.geometric}
\usetikzlibrary{angles,quotes}
\usetikzlibrary{fit}
\usepgfplotslibrary{groupplots}

\colorlet{veccol}{green!50!black}
\colorlet{projcol}{blue!70!black}
\colorlet{myblue}{blue!80!black}
\colorlet{myred}{red!90!black}
\colorlet{mydarkblue}{blue!50!black}
\definecolor{green(pigment)}{rgb}{0.0, 0.65, 0.31}

\tikzset{>=latex} % for LaTeX arrow head
\tikzstyle{proj}=[projcol!80,line width=0.08] %very thin
\tikzstyle{area}=[draw=veccol,fill=veccol!80,fill opacity=0.6]
\tikzstyle{vector}=[-stealth,myblue,thick,line cap=round]
\tikzstyle{unit vector}=[->,veccol,thick,line cap=round]
\tikzstyle{dark unit vector}=[unit vector,veccol!70!black]
\usetikzlibrary{angles,quotes} % for pic (angle labels)
\newsavebox{\myimage}

\definecolor{cadmiumgreen}{rgb}{0.0, 0.42, 0.24}
\definecolor{cobalt}{rgb}{0.0, 0.28, 0.67}

\newcommand{\blu}[1]{{\textcolor{cobalt}{#1}}}

\newcommand{\ist}{\hspace*{.3mm}}
\newcommand{\rmv}{\hspace*{-.3mm}}
\newcommand{\iist}{\hspace*{1mm}}
\newcommand{\rrmv}{\hspace*{-1mm}}
\newcommand{\nn}{\nonumber}

\newcommand{\rv}[1]{\msf{#1}}
\newcommand{\RV}[1]{\bm{\msf{#1}}}

\newcommand{\V}[1]{\bm{#1}}

\DeclareMathAlphabet{\mathsfbr}{OT1}{cmss}{m}{n}%for math sans serif (cmss)
\SetMathAlphabet{\mathsfbr}{bold}{OT1}{cmss}{bx}{n}%for math sans serif (cmss)
\DeclareRobustCommand{\msf}[1]{%
	\ifcat\noexpand#1\relax\msfgreek{#1}\else\mathsfbr{#1}\fi%for math sans serif (cmss)
}
\IEEEoverridecommandlockouts

\usepackage[skip=6pt]{caption}

\begin{document}
\bstctlcite{IEEEexample:BSTcontrol}

% \title{Distributed and Scalable Message-Passing Localization Algorithm and Architecture using D-MIMO}
\title{Low-latency D-MIMO Localization using Distributed Scalable Message-Passing Algorithm}
\vspace{-1mm}
\author{
\IEEEauthorblockN{Dumitra Iancu\IEEEauthorrefmark{1}, 
Liang Liu\IEEEauthorrefmark{1}, 
Ove Edfors\IEEEauthorrefmark{1}, 
Erik Leitinger\IEEEauthorrefmark{2}, 
Xuhong Li\IEEEauthorrefmark{1}\IEEEauthorrefmark{3} }
\\ \vspace{-2mm}
\IEEEauthorblockA{\IEEEauthorrefmark{1} Departament of Electrical and Information Technology, Lund University, Sweden}
\IEEEauthorblockA{\IEEEauthorrefmark{2} Institute of Communication Networks and Satellite Communications, Graz University of Technology, Austria}
\IEEEauthorblockA{\IEEEauthorrefmark{3} Department of Electrical and Computer Engineering, University of California San Diego, USA}
Email: \{firstname.lastname\}@eit.lth.se, @tugraz.at 
\thanks{ This work is funded by the Swedish Foundation for Strategic Research (SSF) project Large Intelligent Surfaces – Architecture and Hardware, and by the Knut and Alice Wallenberg Foundation.}}
 
\IEEEaftertitletext{\vspace{-6mm}}
\maketitle

\begin{abstract}
    Distributed MIMO and integrated sensing and communication are expected to be key technologies in future wireless systems, enabling reliable, low-latency communication and accurate localization. Dedicated localization solutions must support distributed architecture, provide scalability across different system configurations and meet strict latency requirements. We present a scalable message-passing localization method and architecture co-designed for a panel-based distributed MIMO system and network topology, in which interconnected units operate without centralized processing. This method jointly detects line-of-sight paths to distributed units from multipath measurements in dynamic scenarios, localizes the agent, and achieves very low latency. Additionally, we introduce a cycle-accurate system latency model based on implemented FPGA operations, and show important insights into processing latency and hardware utilization and system-level trade-offs. We compare our method to a multipath-based localization method and show that it can achieve similar localization performance, with wide enough distribution of array elements, while offering lower latency and computational complexity.

\end{abstract}

\glsresetall
\IEEEpeerreviewmaketitle
\vspace{\sectionshift}

%--------include sections-----------%
% \vspace{3mm}
\section{Introduction}
\label{sec:Introduction}
6G is expected to feature advancements such as larger signal bandwidth and array aperture, a more decentralized network built on \gls{dmimo} \cite{GustavssonJOM2021}, and support centimeter-level localization, millisecond-level latency and high data rates, across diverse use-case scenarios \cite{BehravanVTM2023}. Localization solutions based on \gls{los} paths or additional \glspl{mpc} both benefit from greatly enhanced spatial resolution, which improves the resolvability of \glspl{mpc}. In general, the former offers lower complexity, while the latter provides higher robustness and accuracy in challenging scenarios (e.g., urban and indoors) with severe multipath, \gls{olos}, and highly dynamic channel conditions. Designing localization solutions that are both computationally efficient, scalable and reliable across diverse system configurations and harsh channel conditions remains a critical and challenging task.

\Gls{bp}-based methods work by passing ``messages'' along the edges of a factor graph that represents the underlying statistical model of the inference problem and provides a highly efficient and scalable solution for multipath-based localization, and has shown state-of-the-art performance in various experiments \cite{LeitingerICLGNSS2016, Florian_Proceeding2018, Erik_SLAM_TWC2019, XuhongICC2024, AlexTWC2024}. The complexity of such methods typically scales linearly with the number of edges in the factor graph, making them efficient for distributed networks but potentially expensive in scenarios involving a large number of distributed units, propagation paths, and highly dynamic channel conditions. Complexity and run-time analyses for several such algorithms have been conducted, providing preliminary estimates of the expected latency \cite{AlexTWC2024, Erik_SLAM_TWC2019}. However, these estimates are dependent on the processor architecture and compiler, and they exceed the latency limits for the expected 6G use-cases, such as collaborative robots \cite{BehravanVTM2023} ($1\rmv-\rmv10$\,ms) or XR-AR applications \cite{HazarikaSensors2023} ($1\rmv-\rmv5$\,ms). Ultimately, an efficient localization solution, benefits from the co-design of algorithm, topology and hardware implementation as it was shown, for example, in the case of deep neural networks\cite{StuderJSAC2025}.

\begin{figure}[!t]
    \centering 
        \includegraphics[width=0.95\columnwidth]{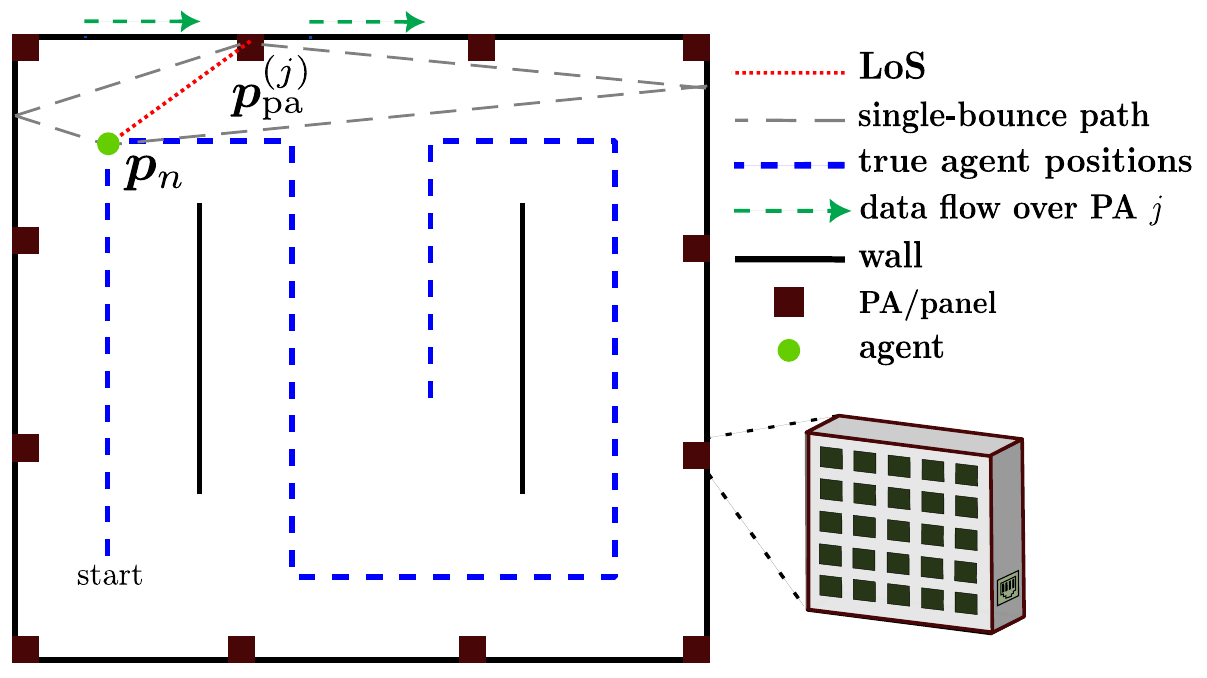}
      %  \vspace{\captionshift}
    \caption{Depiction of an indoor \gls{dmimo} scenario with distributed physical anchors (PAs) of known positions $\V{p}_{\mathrm{pa}}^{(j)}$. PAs are equipped with antenna arrays and local processing capabilities, and the data flow between them follows a daisy-chain topology. At each time $n$, the agent at unknown position $\V{p}_{n}$ transmits radio frequency (RF) signals that reach the PAs via \gls{los} and single-bounce paths reflected off the walls. 
    %The \gls{los} paths can be obstructed by the middle walls.
    }
    \label{fig:pos}
\end{figure}
In this paper, we introduce a low-complexity \gls{bp}-based localization method exploiting \gls{los} paths and adapted to panelized \gls{dmimo} topologies. The main contributions are summarized as follows: (i) The distributed algorithm performs local processing with minimal inter-panel communication, unlike some centralized algorithms that may incur high latency due to centralized data aggregation and processing. Essentially identical operations at each panel facilitate system scalability without hardware redesign when adding more antenna panels; (ii) we formulate a hardware-assisted latency model based on the \gls{fpga} implementation of the building processing blocks of the algorithm, allowing us to make system-level trade-offs and to achieve a more accurate estimate of the potential latency for a real deployment case; (iii) we apply the algorithm to a dynamic indoor scenario with \gls{olos} conditions and show the trade-offs between different key design parameters, such as array size, number of panels and particle number together with latency, that can serve as input for further optimizing the algorithm and hardware implementation through algorithm-hardware co-design. Additionally, we use a multipath-based method \cite{LeitingerICLGNSS2016, Erik_SLAM_TWC2019} as a performance benchmark and demonstrate that the \gls{los}-based approach achieves comparable accuracy as the number of panels increases. This shows that the system can trade a higher panel count for reduced algorithmic processing complexity, without much latency overhead.

% The model can serve as an aid for fast design-space exploration for future testbed implementations and it strengthens the theoretical assumptions by revealing the bottlenecks from a latency and hardware-resource perspective; 
% \subfile{./InputFiles/Introduction} 

\section{Problem formulation and System Model}
\label{sec:SystemModel}
%\subsection{Scenario and problem formulation}
%  The panels are equipped with local processing units and are interconnected, receiving and transmitting data from each other.  processing capabilities The concept is illustrated in Fig.~\ref{fig:pos}
% \vspace{3mm}
We consider a distributed panelized \gls{mimo} system as shown in Fig.~\ref{fig:pos}, consisting of $J$ panels (i.e., \glspl{pa}) at positions $\V{p}_{\mathrm{pa}}^{(j)}$. Each \gls{pa} is equipped with an $N_{\mathrm{a}}$-element antenna array with known orientation. \glspl{pa} are equipped with local processing units and are interconnected via a communication backbone, enabling data exchange between them. The single-antenna mobile agent has an unknown and time-varying position $\V{p}_{n}$. We assume perfect synchronization between panels and the agent. At each time $n$, the agent transmits \gls{rf} signals that are received by the $j$th \gls{pa} via \gls{los} and \gls{nlos} propagation paths such as specular reflections. Reflected paths are typically modeled using \glspl{va} denoting the mirror images of \glspl{pa} on reflective surfaces. The numbers and positions of \glspl{pa} and \glspl{va} are considered known, but their visibilities from the current position $\V{p}_{n}$ are unknown as some paths may be obstructed by objects, e.g., the middle walls shown in Fig.~\ref{fig:pos}.

At each time $n$, as a pre-processing stage, a super-resolution channel estimation algorithm is applied to the observed \gls{rf} signals at each \gls{pa} $j$, providing estimated parameters of $M_{n}^{(j)}$ \glspl{mpc} stacked into the vector $\V{z}_{n}^{(j)} \rmv\triangleq\rmv [\V{z}_{1,n}^{(j) \mathrm{T}} \cdots \V{z}_{M_{n}^{(j)},n}^{(j) \mathrm{T}}]^{\mathrm{T}} \rmv\in\rmv \mathbb{R}^{3M_{n}^{(j)}\rmv\times\rmv1} $, with each $ \V{z}_{m,n}^{(j)} \rmv\triangleq\rmv [d_{m,n}^{(j)}, \iist \varphi_{m,n}^{(j)}, \iist u_{m,n}^{(j)}]^{\mathrm{T}},\, m \in \{1,\dots, M_n^{(j)}\}$ comprising the distance $d_{m,n}^{(j)}$, the \gls{aoa} $\varphi_{m,n}^{(j)}$ and the normalized amplitude $u_{m,n}^{(j)}$ denoting the square root of the component \gls{snr}. Note that $M_{n}^{(j)}$ may differ from the true number of visible paths and is time-varying, and measurement impairments (\glspl{fa} or missed detections) may exist. We further define the vector $\V{z}_{n} \rmv\triangleq\rmv [\V{z}_{n}^{(1) \mathrm{T}} \cdots \V{z}_{n}^{(J) \mathrm{T}}]^{\mathrm{T}} $ stacking measurements from all \glspl{pa}. Using noisy measurements $\V{z}_{n}^{(j)} $, the goal is to sequentially localize the agent exploiting \gls{los} paths to visible \glspl{pa}, leading to a joint problem of agent state estimation and \gls{los} path existence detection.%, as described in the following. 

% $\V{x}_{n} \triangleq [\V{p}_{n}^{\mathrm{T}}, \V{v}_{n}^{\mathrm{T}}]^{\mathrm{T}} $ consisting the position $\V{p}_{n} =[p_{\mathrm{x},n} \iist p_{\mathrm{y},n}]^{\mathrm{T}}$ and the velocity $ \V{v}_{n} =[v_{\mathrm{x},n} \iist v_{\mathrm{y},n}]^{\mathrm{T}} $.  At each discrete time $n$, 

\subsection{System Model}
At each time $ n $, the state of mobile agent is given by $ \RV{x}_{n} \triangleq [\RV{p}_{n}^{\mathrm{T}} \iist \RV{v}_{n}^{\mathrm{T}}]^{\mathrm{T}}$ consisting of the position $ \RV{p}_{n} $ and the velocity $ \RV{v}_{n} =[\rv{v}_{\mathrm{x},n} \iist \rv{v}_{\mathrm{y},n}]^{\mathrm{T}} $. To account for the unknown and time-varying \gls{los} propagation conditions, we introduce for each \gls{pa} the state $\RV{y}_{n}^{(j)} \triangleq [\rv{u}_{n}^{(j)}, \rv{r}_{n}^{(j)}]^{\mathrm{T}}$ with $\rv{u}_{n}^{(j)}$ and $\rv{r}_{n}^{(j)} \in \{0,1\}$ denoting the normalized amplitude and the binary random variable indicating the \gls{los} existence. Specifically, the \gls{los} exists if $r_n^{(j)}=1$, $0$ otherwise. The use of amplitude information enables adaptive detection of \gls{los} paths and helps to capture measurement uncertainties \cite{XuhongTWC2022}. Measurements are subject to data association uncertainties, thus it is not known if the measurement $\V{z}_{m,n}^{(j)} $ originated from the \gls{los}, or it is due to a reflection or a \glspl{fa}. The associations between measurements and \glspl{los}, according to the \gls{pda} \cite{AlexTWC2024},  are described by the association vector $ \RV{a}_{n} \triangleq [\rv{a}_{n}^{(1)} \ist \cdots \ist \rv{a}_{n}^{(J)}]^{\mathrm{T}} $ with entries $ \rv{a}_{n}^{(j)} \triangleq m \rmv\in\rmv \{1,\dots, \rv{M}_n^{(j)}\}$ if $\V{z}_{m,n}^{(j)} $ is a \gls{los} measurement, and $\rv{a}_{n}^{(j)} \triangleq 0$ otherwise.

\section{Distributed algorithm and architecture}
\label{sec:Distributedalgorithmandarchitecture}
%We design our algorithm following a daisy-chain topology, advantageous for its simplicity and scalability\cite{SanchezTSP2020}, starting from centralized \gls{bp} localization algorithms \cite{Erik_SLAM_TWC2019, XuhongTWC2022, AlexTWC2024}, also known as sum-product message passing, to obtain a simpler, distributed algorithm that can perform estimation locally, at each panel. 
In this section, we introduce the \gls{los}-based localization algorithm and the processing architecture for a panelized \gls{dmimo} system following a daisy-chain topology.
%The data exchange between panels scales with the number of particles and
The architecture that we explore in this work is advantageous for its simplicity, modularity and scalability\cite{SanchezTSP2020} from an implementation point of view, as adding or removing one panel does not require any hardware redesigning.
\subsection{Sum-Product Algorithm}
\label{sec:spa}
The localization problem can be summarized as a joint Bayesian inference process on states $\RV{x}_{n} $, $\RV{y}_{n}^{(j)}$, and $\RV{a}_{n} $ given observed (thus fixed) measurements $\V{z}_{1:n}$ of all \glspl{pa} and all times up to $n$. The \gls{los} to \gls{pa} $j$ is claimed to be detected if the marginal posterior existence probabilities $ p(r_{n}^{(j)} = 1 | \V{z}_{1:n}) \rmv>\rmv p_\mathrm{de} $, with $ p_\mathrm{de} $ denoting the detection threshold. 
% The state estimation of agent and detected \glspl{pa} rely on the marginal posterior \glspl{pdf} $f(\V{x}_{n}|\V{z}_{1:n})$ and $f( u_{n}^{(j)} |r_{n}^{(j)}=1, \V{z}_{1:n})$ and are performed by means of the \gls{mmse} estimator \cite{Kay_EstimationTheory}, yields
\Gls{mmse} estimation \cite{Kay_EstimationTheory} of agent state and detected \glspl{pa} are calculated as conditional means
\begin{align}	
    \hat{\V{x}}_{n} &\triangleq \int \V{x}_{n} f(\V{x}_{n}|\V{z}_{1:n})\mathrm{d} \V{x}_{n},
    \label{eq:MMSE_x} \\
    \hat{u}_{n}^{(j)} &\triangleq \int u_{n}^{(j)} f( u_{n}^{(j)} |r_{n}^{(j)}=1, \V{z}_{1:n}) \mathrm{d} u_{n}^{(j)}.
    \label{eq:MMSE_PA} 
    \vspace{0mm}
\end{align}
relying on the marginal posterior \glspl{pdf} $f(\V{x}_{n}|\V{z}_{1:n})$ and $f( u_{n}^{(j)} |r_{n}^{(j)}=1, \V{z}_{1:n})$.
\begin{figure*}[t]
    \centering
    \scalebox{0.96}{
    \hspace{-0mm}\includegraphics{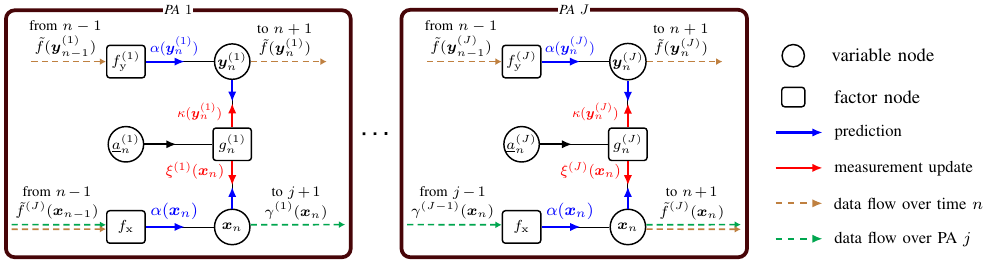}}
    \caption{Factor graph representation of the joint posterior \gls{pdf} \eqref{eq:jointPDF}. The following short notations are used: the state-transition \gls{pdf} for the agent state $f_{\mathrm{x}} = f({\V{x}}_{n}|\V{x}_{n-1})$ for $j=1$, and $ f(\V{x}_{n}^{(j)}|\V{x}_{n-1}^{(j)}) $ for $j >1$, and state-transition for the $j$th \gls{pa} state $f_{\mathrm{y}}^{(j)} = f({\V{y}}_{n}^{(j)}|\V{y}_{n-1}^{(j)})$, the pseudo-likelihood function $g_n^{(j)} = g(\V{x}_{n}, \V{y}_{n}^{(j)}, \V{a}_{n}^{(j)}; \V{z}_{n})$. } 
    \label{fig:FG}
\end{figure*}
Since the direct marginalization of the joint posterior \gls{pdf} $f(\V{x}_{0:n}, {\V{y}}_{0:n}, {\V{a}}_{0:n}|\V{z}_{0:n})$ is computationally infeasible, 
we perform message passing using the \gls{spa} rules on the factor graph in Fig.~\ref{fig:FG} representing the factorized joint posterior \gls{pdf} \eqref{eq:jointPDF}, which efficiently obtains the beliefs $\tilde{f}^{(j)}(\V{x}_{n})$ and $f(\V{y}_{n}^{(j)}|\V{z}_{1:n})$ approximating these marginal posterior \glspl{pdf}. The factorized joint posterior \gls{pdf} is given by
\vspace{-3mm}
\begin{align}
    & \hspace*{0mm}f(\V{x}_{0:n}, {\V{y}}_{0:n}, {\V{a}}_{0:n}|\V{z}_{0:n}) \label{eq:jointPDF} \\
    &\hspace*{1mm} = f(\V{x}_{0}) \prod_{j = 1}^{J} f(\V{y}_{0}^{(j)}) 
    \prod_{n' = 1}^{n} f(\V{x}_{n'}|\V{x}_{n'-1})  \nonumber  \\ \vspace{-1mm}
    &\hspace*{5mm}\times \prod_{j' = 1}^{J} f({\V{y}}_{n'}^{(j')}|\V{y}_{n'-1}^{(j')}) g(\V{x}_{n'}, \V{y}_{n'}^{(j')}, \V{a}_{n'}^{(j')}; \V{z}_{n'}) \nn  \\ \vspace{-1mm}
    & \hspace*{5mm} \times \prod_{j'' = 2}^{J} f(\V{x}^{(j'')}_{n'}|\V{x}^{(j''-1)}_{n'}) g(\V{x}_{n'}^{(j'')}, \V{y}_{n'}^{(j'')}, \V{a}_{n'}^{(j)}; \V{z}_{n'}) \nn \vspace{-2mm}
\end{align}
where $f(\V{x}_{0})$ and $f(\V{y}_{0}^{(j)}) $ denote the initial states at the beginning. The agent state $\RV{x}_{n} $ and the \gls{pa} states $\RV{y}_{n}^{(j)}$ are assumed to evolve independently across $ n $ and $ j $ according to state-transition \glspl{pdf} $ f(\V{x}_{n}|\V{x}_{n-1}) $ and $ f(\V{y}_{n}^{(j)}|\V{y}_{n-1}^{(j)}) $, respectively. Considering the sequential incorporation of agent state across \glspl{pa} at each time $ n $, we also define the agent state-transition \gls{pdf} $ f(\V{x}_{n}^{(j)}|\V{x}_{n-1}^{(j)}) $ for $j >1$. The above state-transition \glspl{pdf} and the pseudo-likelihood function $g(\V{x}_{n}, \V{y}_{n}^{(j)}, \V{a}_{n}^{(j)}; \V{z}_{n}) = g(\V{x}_{n}, u_{n}^{(j)}, r_{n}^{(j)}, \V{a}_{n}; \V{z}_{n})$ are formulated in line with \cite{LeitingerICC2019}. According to the generic \gls{spa} rules, the messages and beliefs involved in the factor graph in Fig.~\ref{fig:FG} are obtained as follows.
\vspace{-6mm}
\subsubsection{Prediction} First, a prediction step from time $n-1$ to $n$ is performed for the agent state and all \gls{pa} states. The prediction messages $\alpha(\V{x}_{n})$ and $\alpha(\V{y}_{n}^{(j)}) = \alpha(u_{n}^{(j)}, r_{n}^{(j)})$ are given by
\begin{align}
\alpha(\V{x}_{n}) &= \int \rmv f(\V{x}_{n}|\V{x}_{n-1}) \tilde{f}(\V{x}_{n-1}) \mathrm{d} \V{x}_{n-1}, \label{eq:preTimeAgent} \\[-2mm] 
\alpha(\V{y}_{n}^{(j)}) &= \int \rmv f(\V{y}_{n}^{(j)}|\V{y}_{n-1}^{(j)}) \tilde{f}(\V{y}_{n-1}^{(j)}) \mathrm{d} \V{y}_{n-1}^{(j)}. \label{eq:preTimePA}
\end{align}
For \gls{pa} $j>1$ at time $n$, the prediction message $\alpha^{(j)}(\V{x}_{n})$ for the agent is given as in \cite{Erik_SLAM_TWC2019}, applying a small Gaussian regularization noise to the agent state from the previous \gls{pa}.

\subsubsection{Measurement Update} The message $\xi^{(j)}(\V{x}_{n})$ sent from the factor node $ g_{n}^{(j)}$ in Fig.~\ref{fig:FG} to the agent state is given by
\begin{align}
\xi^{(j)}(\V{x}_{n}) & = \int \rrmv  \sum_{a_{n}^{(j)}=0}^{M_{n}^{(j)}} \alpha(u_{n}^{(j)}, r_{n}^{(j)}=1) \label{eq:MeaUpdateAgent} \vspace{-1mm}\\
& \hspace{2mm} \times g(\V{x}_{n}, u_{n}^{(j)}, r_{n}^{(j)}=1, \V{a}_{n}; \V{z}_{n}) \mathrm{d} u_{n}^{(j)} + \alpha_{n}^{(j)} \nn 
\end{align}
where $\alpha_{n}^{(j)}$ approximates the nonexistent probability of \gls{los} path to \gls{pa} $j$ at time $n$. Accordingly, the messages $\gamma^{(j)}(\V{x}_{n})$ sent to the next \gls{pa} $j+1$ is given as $\gamma^{(j)}(\V{x}_{n}) = \alpha(\V{x}_{n})\xi^{(j)}(\V{x}_{n})$. The message sent from the factor node $ g_{n}^{(j)}$ to the \gls{pa} state $\kappa(\V{{y}}^{(j)}_{n}) = \kappa(u_{n}^{(j)}, r_{n}^{(j)})$ is given by 
\begin{align}
    \kappa(u_{n}^{(j)}, 1) = \int \rrmv \alpha(\V{x}_{n})\sum_{a_{n}^{(j)}=0}^{M_{n}^{(j)}} g(\V{x}_{n}, u_{n}^{(j)}, 1, \V{a}_{n}; \V{z}_{n}) \mathrm{d} \V{x}_{n},
    \label{eq:MeaUpdatePA}
\end{align} 
and $\kappa(u_{n}^{(j)}, r_{n}^{(j)}=0) \triangleq 1$.

\subsubsection{Belief Calculation}The belief $\tilde{f}^{(j)}(\V{x}_{n})$ of the agent state at the final \gls{pa} $J$ approximating $f(\V{x}_{n}|\V{z}_{1:n})$ is given by 
\begin{align}
	\tilde{f}(\V{x}_{n}) = C_{\mathrm{x},n}\alpha(\V{x}_{n}) \xi^{(J)}(\V{x}_{n})\, .
	\label{eq:BeliefAgent}
\end{align}
The belief $\tilde{f}(\V{y}_{n}^{(j)})$ approximating $f(\V{y}_{n}^{(j)}|\V{z}_{1:n})$ at \gls{pa} $j$ equals %is given by 
\begin{align}
	\tilde{f}(\V{y}_{n}^{(j)}) = C_{\mathrm{y},n}\alpha(\V{y}_{n}^{(j)}) \kappa(\V{{y}}^{(j)}_{n})\, .
	\label{eq:BeliefAgent}
\end{align}
The normalization constants $C_{\mathrm{x},n}$ and $C_{\mathrm{y},n}$ ensure that the beliefs are valid probability distributions. A particle-based implementation is used to efficiently calculate the marginal \glspl{pdf}. We use a ``stacked state'' implementation so the complexity scales linearly in the number of particles and measurements. Regularization and resampling are performed to prevent weight degeneracy.

\subsection{Processing architecture}
We consider a multi-\gls{fpga} based hardware architecture, where each \gls{fpga}-based platform represents a \gls{pa} -- a panel with $N_{\mathrm{a}}$ array elements, \gls{rf} chains and processing capabilities. %\Glspl{fpga} offer several advantages over using \glspl{cpu} or \glspl{asic}, including short prototyping time for dedicated hardware and high flexibility, as they can be easily reconfigured and adapted to various topologies.
The advantage of distributing the compute workload onto panels with processing capabilities is avoiding sending the data from all the antenna panels to a central processing unit, needing to aggregate and process the data jointly, which scales with both the number of particles, as well as the number of panels. With a growing array element number $N_{\mathrm{a}}$ and panel number $J$, in \gls{dmimo}, the centralized processing becomes unscalable from both a latency and a hardware resource perspective. 
 % The platform gives a good eavluaton compred to other platforms
Each \gls{fpga} is directly connected to each other sequentially, in a daisy-chain fashion via Ethernet ports, avoiding any additional router logic that might be needed in case of other topologies.  All the operations described previously and shown in Fig. \ref{fig:FG} for each \gls{pa} are executed locally, on each \gls{fpga}. The message $\gamma^{(j)}(\V{x}_{n})$ from panel $j$ representing the updated agent particle distribution, is transmitted to panel $j+1$, where it serves as the prior distribution for the subsequent update.
This feature of the algorithm eliminates the need for a separate data aggregation stage, as the information from both preceding and current panels is inherently embedded within the current panel's message.
% Once computed, the agent messages $\gamma^{(j)}(\V{x}_{n})$  are passed from panel to panel, thus each panel having to wait for the previous one, whereas 
Once the \gls{pa} beliefs $\tilde{f}(\V{y}_{n}^{(j)})$ are updated, the prediction  $f({\V{y}}_{n}^{(j)}|\V{y}_{n-1}^{(j)})$ can start computing immediately for time step $n+1$, as it is performed independently, in parallel, at each panel. Once the $\gamma^{(j)}(\V{x}_{n})$ reaches the last panel and is updated, we obtain the estimates $\hat{\V{x}}_{n}$. 
%Intuitively, the sequential processing might lead one to expect higher processing latency than a centralized approach.
The next section breaks down the numbers of the time it takes for one inference, in one time step,  based on implemented hardware blocks.

\section{FPGA Implementation and Latency Analysis}
\label{sec:LatencyAnalysis}

\begin{figure}[!t]
    \centering
    \scalebox{0.95}{\includegraphics{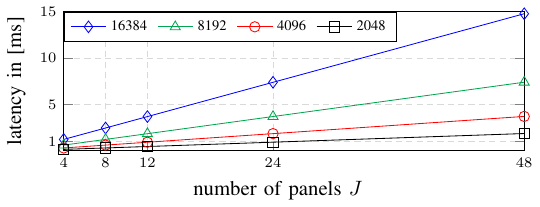}}
    %\vspace{-5mm}
    \caption{Latency of the panelized \gls{dmimo} localization system, based on Eq.~\ref{eq:latency} and Table~\ref{table:latency} for varying particle numbers $N_{\mathrm{p}}$, illustrated with different markers and colors, and a fixed number of measurements, $M_n^{(j)} = 6$.}	 
    \label{fig:latency}
\end{figure}

We are interested in analyzing the latency achievable for this algorithm in this particular setup, but constrained by available hardware resources. To this end, we developed a cycle-accurate latency model that leverages latency numbers extracted from a \gls{fpga} implementation.
The operations described in the previous section are accelerated using \gls{hls} with Matlab HDL Coder, on the AMD RFSoC ZCU216 platform. ZCU216 integrates powerful DSP capability with 16-channel direct RF sampling, making it an attractive hardware platform for implementing 6G D-MIMO systems. 
The implementation has been done for a number of $N_{\mathrm{P}} =4$ particles in order to utilize as few hardware resources as possible, whilst considering the data dependencies between particles present in the algorithm. Table \ref{table:latency} shows the latency on a clock cycle level, as well as the \gls{fpga} utilization breakdown for the number of \glspl{lut}, flip-flops, and \gls{dsp} slices. To favor low latency, no area optimizations, a minimal number of registers in the datapath and a precision of $32$-bit fixed-point have been chosen. Loop unrolling is considered whenever possible to process 4 particles in parallel. For the panel interconnect latency, we consider 25G Ethernet links between the panels, with one transfer taking $ \tau_\mathrm{fronthaul}\rmv=\rmv0.87 \mu s$ for the full throughput, latency based on an implemented Ethernet IP.
% This is the latency from the first sent data to the first received data. So evenif we cannot fit 16384 particles in a jumbo frame, and we fit the minimum of 174/3 just 
For the considered scenario in Fig. ~\ref{fig:pos}, the distances between panels are small enough to consider the signal propagation through optical fiber negligible. Additional buffering resources due to protocol handshakes is not considered in this analysis, as all implemented blocks are operating at matching throughput. 
\begin{table}[t]
\vspace{5mm}
\caption{Latency and utilization profile for operations described in \ref{sec:Distributedalgorithmandarchitecture}, @200Mhz on RFSoC ZCU216, for $N_\mathrm{P}\rmv=\rmv4$. The operation in \blu{blue}, on row three, is not added to the total latency as it can be computed in parallel with other blocks.} 
 \resizebox{\columnwidth}{!}{
\begin{tabular}{ccccc}
\hline \hline
\multirow{2}{*}{Operation} & \multirow{2}{*}{Latency(cc)} & \multicolumn{3}{c}{Utilization(\%)} \\ 
                                 &                              & LUT        & FF        & DSP        \\ \hline \hline
$f(\V{x}_{n}|\V{x}_{n-1})$, $j=1$         & 3                            &  0.21          &  0.06         &   0.89         \\
$f(\V{x}_{n}|\V{x}_{n-1})$, $j>1$       & 3                            & 0.06           & 0.11          &  -         \\
\blu{$f({\V{y}}_{n}^{(j)}|\V{y}_{n-1}^{(j)})$}        & \blu{3}                           &   \blu{0.04}         &     \blu{0.06}    &   \blu{ - }    \\
$g(\V{x}_{n}, \V{y}_{n}^{(j)}, \V{a}_{n}^{(j)}; \V{z}_{n})$              & 9 + $2(M_n^{(j)}-1) $               & 19.65         &  0.56         &      14.79      \\
$\tilde{f}(\V{y}_{n}^{(j)}) $             & 5                            &    2.77          & 0.17        &  0.38         \\
$\tilde{f}(\V{x}_{n}^{(j)})$           & 3                            & 10.10        & 0.07         & 4.12         \\
total for \gls{pa} $j=1$          & 20 + $2(M_n^{(j)}-1)$                         & 32.77      & 0.92         & 20.18        \\
total for \gls{pa} $j>1$          & 20 + $2(M_n^{(j)}-1)$                         & 32.62     & 0.97         & 19.29       \\
$\hat{\V{x}}_{n}$                 & 3                &  0.03    & 0.04   & -  \\
ethernet transfer\footnote{The reference Ethernet IP works on a different frequency, but it is normalized to $200$MHz for comparison purposes.}                & 174                          &   5.92         &    3.95       &         - \\
\hline \hline
\end{tabular}
}
\label{table:latency}
\end{table}
Based on the daisy-chain topology, we can model the latency in clock cycles for the entire panelized \gls{dmimo} system for the worst case scenario (in terms of latency) when all panels are contributing to the estimate, as follows: 
\begin{equation}
\tau= J \times (20 + 2(M^{(j)}_n-1)) \times \frac{N_\mathrm{P}}{4} + (J-1) \times \tau_\mathrm{fronthaul},
\label{eq:latency}
\end{equation} 
where $J$ is the number of panels in the system, $N_{\mathrm{P}}$ is the number of particles and $M^{(j)}_n$ is the number of measurements acquired at time $n$ and at panel $j$. Based on the utilization metrics shown in Table \ref{table:latency}, we assume that 4 particles can be processed in parallel. For a general case with an arbitrary number of particles $N_{\mathrm{P}}$, we adopt a time-multiplexed processing approach.
%We consider that we can process 4 particles in parallel, as per utilization metrics presented in Table \ref{table:latency}, and for an arbitrary number of particles $N_{\mathrm{P}}$, we are using a time-multiplexed approach.
For the last panel, a number of $\frac{\log_{2}N_{\mathrm{P}}}{2}$ clock cyles is added, for the final computation of $\hat{\V{x}}_{n}$, assuming that an adder tree is used for this operation. 
The prediction of the \gls{pa} state $f({\V{y}}_{n}^{(j)}|\V{y}_{n-1}^{(j)})$, for time $n+1$ is performed in parallel with other tasks, as it can start the computation as soon as the belief $\tilde{f}(\V{y}_{n}^{(j)})$ at \gls{pa} $j$ has finished computing. Thus, this block is not added to the total latency. The data from one measurement is processed in two clock cycles and for simplicity we consider a fixed number of $M_n^{(j)}$ at every panel.
Initialization time for each of the panels' states as well as the latency of the parametric channel estimation are not included in this analysis, as they can be performed in parallel by each individual panels, thus representing a bias in the total computed latency. Fig.~\ref{fig:latency} shows the calculated latency for different panel and particle configurations, based on Eq.~\ref{eq:latency}. 
% A \gls{hls} implementation has the advantage of faster prototyping but it comes at the expense of the fine-grained control one has in the details of the hardware implementation. 
Timing and utilization metrics could vary depending on the hardware design parameters such as number representation or the degree of pipelining, leaving room for lowering the latency even more through various optimizations. % Our analysis together with hardware implementation, which takes around $33$\% of \gls{fpga} resources, opens the possibility for parametric channel estimation and for potential \gls{dmimo} detection algorithm IPs that can be processed in parallel, whilst proving milisecond latency is possible for certain system configurations.
The implementation of the proposed localization algorithm only occupies around $33$\% of \gls{fpga} resources, leaving ample room for implementing parametric channel estimation and other services, such as communication.

From the latency results and model, we can extract the following conclusions: (i) the execution time is dominated by the number of particles as $N_{\mathrm{P}}$ increases, with the number of panels or measurements playing a smaller role in the total system latency. With fewer particles but more panels we can reach ms-level latency; (ii) the main culprit is calculating the likelihood function, which is expected since it uses divisions, trigonometric and mathematical functions such as \gls{erfc} that require specialized \glspl{lut} or approximate functions that typically use \gls{dsp} slices. Moreover, the latency is dependent on the number of measurements $M_n^{(j)}$ at time $n$. This can be partially solved by processing multiple measurements in parallel at the expense of extra hardware; (iii) the systematic resampling operation used in agent and PA estimation, notorious for its computational complexity, remains a latency bottleneck based on our analysis even with the time-multiplexed approach of processing $4$ particles at a time. This can also be improved at the expense of extra hardware resources, although sacrificing hardware efficiency. 

\section{Performance Evaluation and Trade-offs}
\label{sec:PerformanceAndTradeOff}
\begin{figure}[!t]
    \centering
    \hspace{-3mm}\scalebox{0.95}{\includegraphics{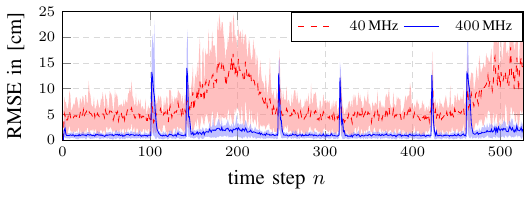}}
    \vspace{-0mm}
    \caption{Agent position \glspl{rmse} averaged over $20$ simulation runs are shown for \textit{Loca-\gls{los}} across time steps, with $N_{\mathrm{P}}=4096$ and $B_{\mathrm{w}}= \{40, 400\}$\,MHz. The shaded bands plotted around the \glspl{rmse} represent the spread of error samples.}	
    \label{fig:modeBelief}
\end{figure}

Localization performance and associated trade-offs are evaluated using synthetic measurements generated for the \gls{dmimo} scenario shown in Fig.~\ref{fig:pos}, which represents a $30\rmv\times\rmv30~\mathrm{m}^2$ indoor area bounded by four outer walls and containing two interior walls resulting in time-varying visibility conditions across different \glspl{pa}, making the simulations more practical and representative of real-world environments. The evaluation focuses on the localization algorithm without involving the channel estimator, therefore noisy \gls{mpc} parameter estimates, used as the measurements $\V{z}_{n}$ for the \gls{spa} algorithm in section~\ref{sec:spa} are directly synthesized, which include components from visible \glspl{los}, single-bounce reflections, and associated impairments manifested like \glspl{fa} and missed detections. We present the results for two different methods: (i) \textit{Loca-LoS}: i.e., the proposed method, localization exploiting \gls{los} paths; (ii) \textit{Loca-MPC}: localization exploiting \glspl{los} and single-bounce paths \cite{LeitingerICLGNSS2016}. The performance is measured in terms of the \glspl{rmse} of the agent position. %For the two setups, the following configurations and simulation parameters are commonly used unless otherwise stated.
% associated impairments manifested

\subsection{Simulation Setup}

%We consider a 30x30$m^2$ room, with two walls in the middle. A one-antenna user transmits a root-raised-cosine pulse with a center frequency of $f_c = 28$GHz and bandwidth values of $B=400$MHz and $B=40$MHz, values validated in \cite{MichielTAP2025} and \cite{ChristianJECNC2023}.
%with two optional bandwidth configurations: $B=400$\,MHz \cite{MichielTAP2025} and $B=40$\,MHz \cite{ChristianJECNC2023}.

The \gls{dmimo} system operates at carrier frequency of $f_{\mathrm{c}} = 28$\,GHz and with bandwidth $B_{\mathrm{w}}$. A $ N_{\mathrm{a}}$-element uniform rectangular array with inter-element spacing of $\lambda/4$ is used at all \glspl{pa} with known orientations of $0^{\circ}$. Over $526$ discrete time steps, visible \gls{los} path and single-bounce paths to each \gls{pa} with time-varying distances, \glspl{aoa}, and normalized amplitudes were synthesized. We performed $20$ simulation runs, and for each simulation run, noisy measurements were generated by adding noises (determined based on the Fisher information) to the true path parameters \cite{Erik_SLAM_TWC2019,XuhongICC2024}, and further stacked together with the generated \gls{fa} measurements. The \glspl{pdf} of random variable states are represented by $N_{\mathrm{p}}$ particles each. \gls{los} path detection threshold is $p_{\mathrm{de}} = 0.5$. In the following, the results are presented for the following design configurations: (i) bandwidth $B_{\mathrm{w}}= \{40, 400\}$\,MHz \cite{ChristianJECNC2023,MichielTAP2025}; (ii) panel number\footnote{Panels are evenly distributed along the four outer walls (shown in Fig.~\ref{fig:pos}). For $J=2$, the two panels at the upper-left and upper-right corners are used.}, $J \in \{2, 4, 8, 12, 24, 48 \} $; (iii) rectangular array size, $ N_{\mathrm{a}} \in \{25, 49, 100, 144, 289 \} $, e.g., $N_{\mathrm{a}}=25$ denotes a $5\rmv\times\rmv5$\,array; (iv) particle number $N_{\mathrm{P}}\in \{2048, 4096, 8192, 16384 \} $. %; (v) simulated SNRs.  

%\cblue{update this part when results are all there. } 

 % a \ac{dmimo} system operating at $f_{\mathrm{c}} = 6$\,GHz with bandwidth of $1$\,GHz is used. The same $ 5 \times 5 $ uniform rectangular array with inter-element spacing of $\lambda/4$ is used at both \acp{pa} and the mobile agent. Over $ 307 $ time steps, propagation paths experiencing up to double bounces with time-varying distances, \acp{aoa}, \acp{aod}, and normalized amplitudes were synthesized. We assume that the \ac{pa} orientations are $0^{\circ}$ and the true agent orientation is consistent with the direction of movement. The amplitude of each path is assumed to follow free-space path loss and additionally attenuated by $3$\,dB after each bounce on a surface. The output \ac{snr} at $1$\,m from the \ac{pa} is assumed to be $\mathrm{SNR}_{\mathrm{1m}}=30$\,dB, according to which the measurement noise standard deviation is further calculated based on the Fisher information \cite{Thomas_Asilomar2018, LeitingerICC2019, XuhongTWC2022}. In each simulation run, noisy measurements are generated according to \eqref{eq:LHF_dist}--\eqref{eq:LHF_normAmp}, i.e., adding noises (determined based on the Fisher information) to the true path parameters, and stacked into the vector $\V{z}_{n}^{(j)}$. In addition, \ac{fa} measurements are generated with mean number $ \mu_{\mathrm{fa}}=2 $ and added to $\V{z}_{n}^{(j)}$.

\begin{figure}[!t]
    \centering
    \vspace{1mm}
    \scalebox{0.95}{\includegraphics{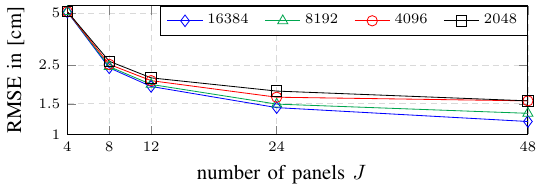}}
    \vspace{-5mm}
    \caption{Agent position \glspl{rmse}, for \textit{Loca-LoS}, averaged over $20$ simulation runs and over time steps, with $B_{\mathrm{w}} = 400$\,MHz, fixed array size $N_{\mathrm{a}}=25$ and varying number of particles $N_\mathrm{p}$.}	 
    \label{fig:perfLoS}
\end{figure}

% \begin{figure}[!tbp]
%   \centering
%   \subfloat[Flower one.]{\scalebox{0.45}{
%     \input{Figures/LoS}
%         }\label{fig:f1}}
%  % \hfill
%   \subfloat[Flower two.]{ \scalebox{0.45}{
%     \input{Figures/LoS}
%         } \label{fig:f2}}
%   \caption{My flowers.}
% \end{figure}

\subsection{Results and Trade-Offs}

\begin{figure}[!t]
    \centering
    \scalebox{0.95}{\includegraphics{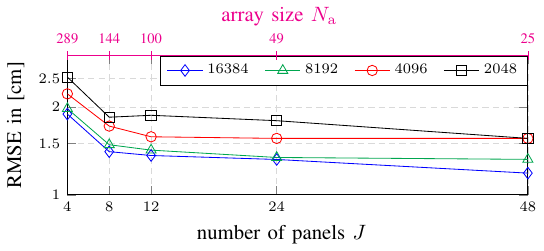}}
    \vspace{-5mm}
    \caption{Agent position \glspl{rmse} for \textit{Loca-LoS} averaged over $20$ simulation runs and over time steps, for $B_{\mathrm{w}} = 400$\,MHz, with a varying number of panels (bottom $x$-axis) and their corresponding array size (top $x$-axis) for different number of particles $N_\mathrm{p}$ denoted by different colors and markers.}	 
    \label{fig:losarraysize}
\end{figure}

\begin{figure}[!t]
    \centering
    \hspace{-2mm}\scalebox{0.95}{\includegraphics{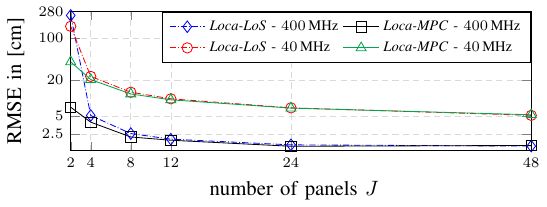}}
    \vspace{-0mm}
    \caption{Agent position \glspl{rmse} are shown for \textit{Loca-\gls{los}} with \textit{Loca-\gls{mpc}}, averaged over $20$ simulation runs and over time steps, with $N_{\mathrm{P}} = 4096$, $B_{\mathrm{w}}= \{40, 400\}$\,MHz and $N_\mathrm{a}=25$.}	
    \label{fig:mpc}
\end{figure}

Given the use-case scenario requirements of, for example, collaborative robots of $1–10$\,ms latency, as expected for 6G systems \cite{BehravanVTM2023}, and the results in Fig. \ref{fig:latency}, we settle on closely analyzing \textit{Loca-\gls{los}} for $N_{\mathrm{p}}=4096$, $J=24$ and $N_\mathrm{a}=25$. Fig.~\ref{fig:modeBelief} shows the \gls{rmse} with error spread of $80\%$ sample-quantile intervals. In both cases, the increased error around steps $200$ and $500$, primarily arises from \gls{los} information being concentrated vertically, as visible \glspl{pa} are placed on the upper and lower walls, leading to greater horizontal position uncertainty. The error peaks observed for $B_{\mathrm{w}} = 400$\,MHz are attributed to the linear near constant-velocity motion model used for the agent state, as well as to the peaky likelihoods and the limited number of particles used which also lead to performance degradation with fewer available measurements, as illustrated in Fig.~\ref{fig:mpc} for panel number $J=2$. 
%Besides, the error peaks are mainly due to the small number of particles used, leading to a less reliable distribution representation, reduced diversity and potentially worse performance with sharp likelihoods (from large bandwidth), fast agent motion (like turns), or less available measurements as illustrated in Fig.~\ref{fig:mpc} for panel number $J=2$. 
Fig.~\ref{fig:perfLoS} shows the \glspl{rmse} averaged over the simulation runs and time steps, for \textit{Loca-\gls{los}} when varying the number of panels, $J$ and the number of particles $N_{\mathrm{p}}$, for a fixed array size of $N_{\mathrm{a}}=25$. %of adding panels with a fixed number of array element $ N_{\mathrm{a}}= 25$ per panel and when we increase the antenna array size, but lower the number of panels. 
The difference in performance is not significant, making the use of less particles an enticing approach from a latency viewpoint. Fig.~\ref{fig:losarraysize} presents an extended analysis by illustrating the performance for various panel number and array size combinations.
As anticipated, larger array size enhances performance, enabling fewer panels to reach cm-level accuracy and benefiting the system in terms of latency, as seen from Fig.~\ref{fig:latency}.
 %The reason one might pick fewer antennas in the antenna array and more panels, over the performance gain, is the potential latency increase in the parametric channel estimation stage.
 No runs diverged, proving that robustness is possible with a smaller number of particles, while keeping the accuracy high with sufficient number of panels.
% \begin{figure}[!t]
% 	\centering
% 	\captionsetup[subfigure]{oneside,margin={0.85cm,0cm}}
% 	\hspace{-2mm}\subfloat[]{\scalebox{0.95}{\input{Figures/Lostrack.tex}}\label{subfig:40}} \\[-3.5mm]
% 	\captionsetup[subfigure]{oneside,margin={0.85cm,0cm}}
% 	\hspace{-2mm}\subfloat[]{\scalebox{0.95}{\input{Figures/Lostrack400.tex}}\label{subfig:400}} 
% 	\caption{Agent position \gls{rmse} and $80\%$ confidence interval for the estimated track in Fig. \ref{fig:pos} for $N_p=4096$ and (a) $B_w = 40$\,MHz, (b) $B_w =400$\,MHz}	 
% 	\label{fig:modeBelief}
% \end{figure}
% \begin{figure}[!t]
%     \centering
%     \captionsetup[subfigure]{oneside,margin={0.85cm,0cm}}
%     \hspace{-10mm}\subfloat[]{\hspace{7mm}\scalebox{0.95}{\input{Figures/LoSfixedtotalantennas}}\label{subfig:1}} \\[-3.5mm]
%     \captionsetup[subfigure]{oneside,margin={0.85cm,0cm}}
%     \hspace{-10mm}\subfloat[]{\hspace{7mm}\scalebox{0.95}{\input{Figures/LoS}}\label{subfig:2}} 
%     \vspace{-2mm}
%     \caption{Agent position \glspl{rmse}, for \textit{Loca-LoS} based on synthetic measurements, averaged over $20$ simulation runs and over time steps, with $B_{\mathrm{w}} = 400$\,MHz. \eqref{subfig:1} illustrates the performance with increasing panels and different array sizes, shown on top and \eqref{subfig:2} for a fixed array size $N_{\mathrm{a}}=25$.}
%     % \vspace*{-2mm}
%     \label{fig:perfLoS}
% \end{figure}

Finally, Fig. \ref{fig:mpc} shows the comparison between \textit{Loca-\gls{los}} and \textit{Loca-\gls{mpc}} for varying panel number $J$ and bandwidth. \textit{Loca-\gls{mpc}} exploits \glspl{mpc} associated to \glspl{va} for agent localization. With perfectly known \gls{va} positions, \textit{Loca-\gls{mpc}} establishes a lower bound for multipath-based localization performance. 
% As expected, $B_{\mathrm{w}}=400$\,MHz provides higher delay resolution, thus outperforms $40$\,MHz for both methods. Regarding the impact of varying the panel number $J$, fewer panels imply severe \gls{olos} and \gls{nlos} conditions and scenarios where deploying multiple anchors is not feasible. In such cases, \textit{Loca}-\gls{mpc} shows greater robustness and reliability than \textit{Loca}-\gls{los}, benefiting from additional geometric information provided by \glspl{va} which enhances spatial diversity. With more panels (e.g., $J\rmv\ge\rmv12$ in our simulation), \textit{Loca}-\gls{los} performs similarly to \textit{Loca-\gls{mpc}}, but offers lower computational complexity and is potentially more suitable for latency-sensitive use cases envisioned in 6G. 
As expected, $B_{\mathrm{w}}=400$\,MHz provides higher delay resolution, thus outperforming $40$\,MHz for both methods. Scenarios with fewer panels where deploying multiple anchors is not feasible, typically result in severe \gls{olos} and \gls{nlos} conditions. In such cases, \textit{Loca-MPC} shows greater robustness and reliability than \textit{Loca-\gls{los}}, benefiting from additional geometric information provided by \glspl{va} which enhances spatial diversity. With more panels (e.g., $J\rmv\ge\rmv12$ in our simulation), \textit{Loca-\gls{los}} performs similarly to \textit{Loca-MPC} and offers lower computational complexity. Moreover, the potential of reaching millisecond-level latency as shown in Fig. \ref{fig:latency}, makes it an  attractive approach for latency-sensitive use cases.

\section{Conclusion}
\label{sec:Conclusion}

We presented a low-complexity \gls{bp}-based localization solution that only leverages \gls{los} paths and is adaptable to D-\gls{mimo} topologies. We have shown that it achieves similar performance as its \gls{mpc}-based counterpart for a wider array distribution, i.e. more array elements and antenna panels. Moreover, we analyzed the latency of the algorithm by providing a cycle-accurate latency model based on \gls{fpga} implementation, proving that milisecond-level latency is achievable.
%attractive from the perspective of using less particles and
% In terms of hardware implementation, \rd{a bit confusing with this sentence} this is just an initial estimate that can be optimized and different design choices can be made to trade-off latency for area. Finer-grained optimization in terms of parallel processing can be done to achieve even lower latency with the same hardware.
% The latency introduced by the channel state estimator can also be explored further, in order to give a better picture of the whole system.
% \Dumitra{General takeaway from making alg-hw co-design more efficient, include interesting papers}
To better understand overall system performance, further improvements can be considered, such as validating the algorithm with real measurement data, accounting for imperfect panel synchronization, and incorporating the latency introduced by parametric channel state estimation. The latter introduces new trade-offs, as larger array sizes or bandwidths may influence overall latency.

 % However, larger array sizes may increase the latency of parametric channel estimation, potentially making a configuration with more panels and fewer array elements preferable.
% There are still challenges and potential improvements that consider the more practical side of things from the algorithm perspective that should be addressed, among them being validating the algorithm with real measurement data or taking the imperfect synchronization between the panels into account. Furthermore, adding the latency introduced by the parametric channel state estimation to the analysis can provide a better overview of the whole system.
%The extraction on the \gls{los} components from cluttered measurements can alse be refined 

\section{Acknowledgements}
The authors would like to thank Lina Tinnerberg for providing the FPGA implementation of the Ethernet IP.

\balance
\bibliographystyle{IEEEtran}
\bibliography{references}

\end{document}